\documentclass[12pt,a4paper,final]{iopart}

\usepackage{iopams}  
\usepackage{graphicx}
\usepackage{bbm}
\usepackage{color}

\begin{document}

\title[Tuning the motility and directionality of self-propelled colloids]{Tuning the motility and directionality of self-propelled colloids}

\author{Juan Ruben Gomez-Solano$^{1,*}$, Sela Samin$^3$, Celia Lozano$^{1,2,*}$, Pablo Ruedas-Batuecas$^1$, Ren\'e van Roij$^3$, and Clemens Bechinger$^{1,2,*}$}

\address{$^1$2. Physikalisches Institut, Universit\"at Stuttgart, Pfaffenwaldring 57, 70569 Stuttgart, Germany}
\address{$^2$Max-Planck-Institute for Intelligent Systems, Heisenbergstrasse 3, 70569 Stuttgart, Germany}
\address{$^3$Institute for Theoretical Physics, Center for Extreme Matter and Emergent Phenomena, Utrecht University, Princetonplein 5, 3584 CC Utrecht, The Netherlands}

\address{*Current address: Fachbereich Physik, Universit\"at Konstanz, Konstanz D-78457, Germany}

\eads{\mailto{ruben.gomez-solano@uni-konstanz.de}}

\begin{abstract}
Microorganisms are able to overcome the thermal randomness of their surroundings by harvesting energy to navigate in viscous fluid environments. In a similar manner, synthetic colloidal microswimmers are capable of mimicking complex biolocomotion by means of simple self-propulsion mechanisms. Although experimentally the speed of active particles can be controlled by e.g. self-generated chemical and thermal gradients, an {\it{in-situ}} change of swimming direction remains a challenge. In this work, we study self-propulsion of half-coated spherical colloids in critical binary mixtures and show that the coupling of local body forces, induced by laser illumination, and the wetting properties of the colloid, can be used to finely tune both the colloid's swimming speed and its directionality. We experimentally and numerically demonstrate that the direction of motion can be reversibly switched by means of the size and shape of the droplet(s) nucleated around the colloid, depending on the particle radius and the fluid's ambient temperature. Moreover, the aforementioned features enable the possibility to realize both negative and positive phototaxis in light intensity gradients. Our results can be extended to other types of half-coated microswimmers, provided that both of their hemispheres are selectively made active but with distinct physical properties.
\end{abstract}

\section*{Introduction}

Locomotion at the microscale is an intriguing non-equilibrium phenomenon that has recently attracted a great deal of attention in soft matter physics~\cite{Romanczuk:2012,Elgeti:2015,Bechinger:2016} and applied science~\cite{Patra:2013,Ahmed:2014}. In the natural realm, microorganisms, such as bacteria and algal cells, are capable to propel themselves through  viscous liquids despite the inertialess flows they create and the random thermal collisions with the surrounding fluid molecules. Their motility is achieved by means of internal biochemical processes that allow them to break the time-reversibility at extremely low Reynolds numbers, e.g. by performing non-reciprocal flagellar beating and rotation, thus converting energy into directed motion with a well-defined polarity~\cite{Purcell:1977}. Moreover, many of these microorganisms, e.g. {\emph{Myxococcus xanthus}}~\cite{Wu:2009}, {\emph{Pseudomonas putida}}~\cite{Theves:2015}, {\emph{Pseudoalteromonas haloplanktis}}, and {\emph{Shewanella putrefaciens}}~\cite{Barbara:2003}, are also able to totally reverse their direction of motion with respect to their main axis, which, depending on the specific environmental conditions, allows them to move forward or backward. This directional reversal results in unexpected dynamical behavior~\cite{Grossman:2016}, such as a large diffusive spreading under geometrical confinement~\cite{Theves:2015} and the formation of traveling wave patterns in dense bacterial colonies~\cite{Boerner:2002}.

In recent years, the first generation of synthetic micro- and nano-swimmers has been developed in order to emulate complex swimming strategies based on well-controlled  physicochemical processes. For instance, actuated swimmers are able to perform a directed motion by controlling their position and/or their orientation by externally applied fields, e.g. magnetic~\cite{Tierno:2008,Gosh:2009}, acoustic~\cite{Wang:2012} and optical~\cite{Dai:2016}. 
On the other hand, active Janus colloids with a symmetry axis determined by two chemically-distinct surfaces can undergo active Brownian motion in liquids, similar to the run-and-tumble motion of {\emph{Escherichia coli}}~\cite{Cates:2013}. In such a case, a synthetic microswimmer is able to autonomously achieve directed motion along its symmetry axis by self-generated chemical~\cite{Moran:2017} or thermal~\cite{Kroy:2016} gradients, which create slip flows on its surface~\cite{Anderson:1989,Golestanian:2007}. For example, catalytic colloids made of inert polystyrene or SiO$_2$ and with a partial active Pt-coating can self-propel in aqueous H$_2$O$_2$ solutions by diffusiophoresis either toward or away from the active site~\cite{Howse:2007,Ke:2010,Palacci:2010}, where the swimming direction can be strongly affected by the detailed particle shape~\cite{Michelin:2017} and ionic effects in the solvent~\cite{Brown:2014}. Although the particle orientation is in turn randomized by rotational diffusion, its directionality remains constant~\cite{Ebbens:2011}, i.e. the propulsion velocity is always either parallel or anti-parallel  to the  particle orientation. For such synthetic microswimmers, the speed can be varied by, e.g., the bulk H$_2$O$_2$ concentration~\cite{Howse:2007} and the active site coverage~\cite{Popescu:2010,Nourhani:2016}, and depends on the particle size~\cite{Ebbens:2012,Brown:2017}. A similar behaviour is observed for active thermophoretic colloids, whose directionality is determined by the sign of the particle's Soret coefficient~\cite{Jiang:2010}. Despite their ability to exhibit finely-tunable motility, a directional reversal akin to that of natural microswimmers has not yet been experimentally demonstrated for active colloids. Apart from being a common feature in nature, this is also a desirable attribute for potential biomedical applications, e.g., drug delivery and tissue engineering, where the swimming direction can be readily switched depending on specific tasks for cargo transport and sorting in complex environments~\cite{Nelson:2010}.

In this work, we report on the accurate {\it{in-situ}} tuning of both the speed and directionality of active two-faced spherical colloids in binary mixtures. The active particle velocity is determined by the size and shape of one or two single-phase droplet(s) that nucleate around the particle, which is induced by means of laser-heating of the colloid's asymmetrically light-absorbing surface. The anisotropy of the droplet shape exerts a net body force, which in turn leads to the particle self-propulsion, and that in contrast to catalytic and thermophoretic active colloids cannot be accounted for by a slip-velocity \cite{Samin:2015}. We find that the propulsion speed is a non-monotonic function of the applied laser intensity. While it linearly increases at sufficiently low intensities and is independent of the particle size, it sharply reverses its direction above a certain intensity threshold when a second droplet covers the uncapped surface. In agreement with numerical calculations, we demonstrate that such an intensity threashold linearly depends on the inverse of the particle radius and on the environmental temperature. Therefore, this non-monotonic dependence of the propulsion speed enables a change in the swimming directionality that can be reversibly varied by means of laser illumination. Remarkably, these unique features allows us to experimentally realize in a straightforward manner both positive and negative phototaxis, i.e. the ability of these synthetic microswimmers to sense a light gradient and to move toward or away from it, respectively.

\section*{Results}

\begin{figure}[ht]
	\includegraphics[width=\linewidth]{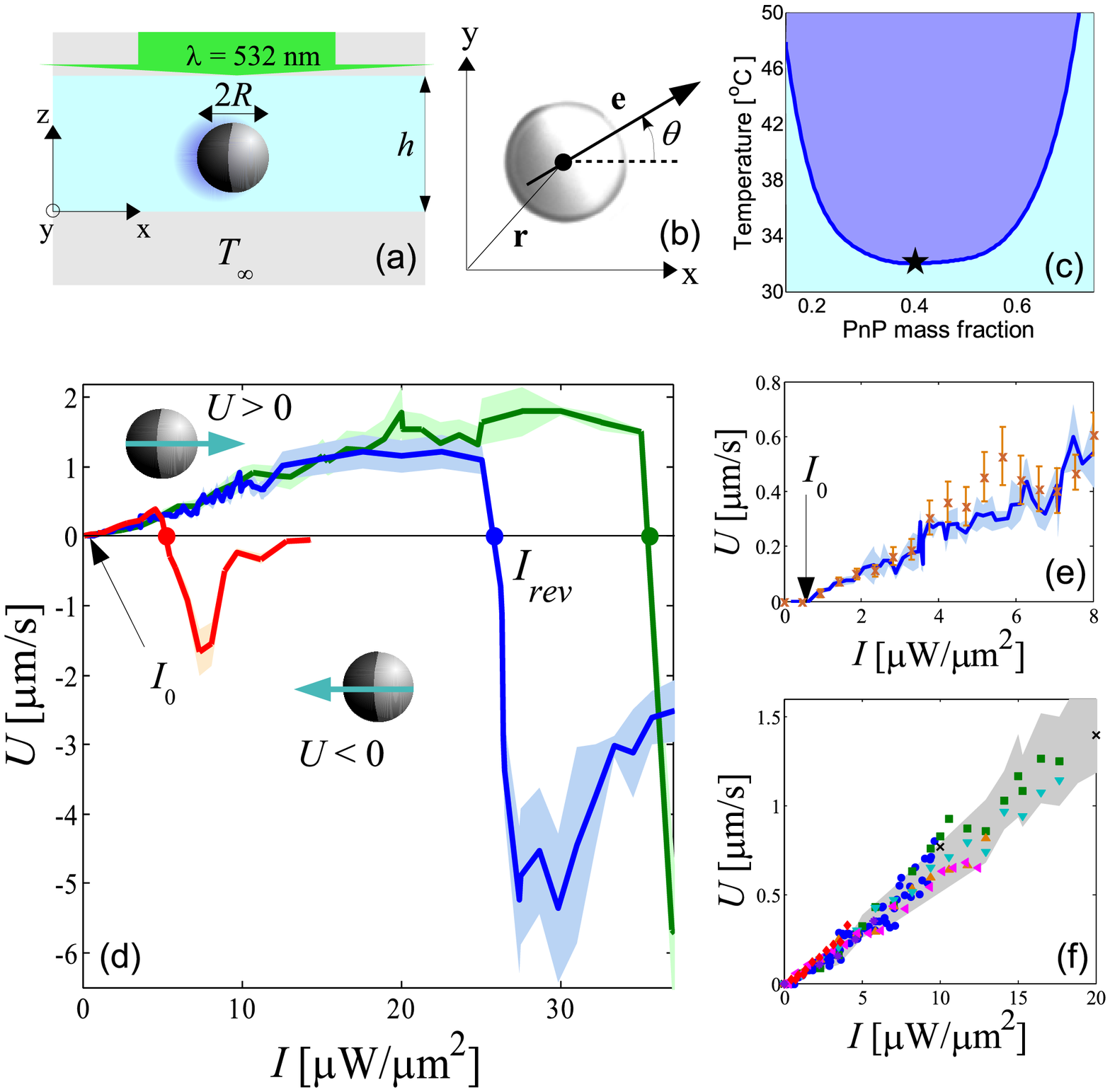}
	\caption{(a) Schematic representation of the experimental setup to induce self-propulsion of a half-coated particle suspended in a binary mixture. (b) Sketch of the 2D particle's position $\mathbf{r}$ and orientation $\mathbf{e}$ on the x-y plane of Fig.~\ref{fig:fig1}(a). (c) Phase diagram of the binary mixture of PnP and water. The darker area above the solid line represents the two-phase region, while the black star corresponds to the critical point ($\phi_c =0.4$. $T_c = 31.9^{\circ}$~C). (d) Dependence of the propulsion speed $U$ as a function of the heating intensity $I$ for particles of radius $R= 3.14\,\mu\mathrm{m}$ (green dashed line), $R= 3.88\,\mu\mathrm{m}$ (blue solid line), and $R= 11.78\,\mu\mathrm{m}$ (red dotted-dashed line) at bath temperature $T_{\infty} = 25^{\circ}$~C. The shaded areas represent the corresponding errors bars, computed over different particles in the same sample. (e) Propulsion speed at $T_{\infty} = 25^{\circ}$~C of $R= 3.88\,\mu\mathrm{m}$  particles with carbon caps of different thickness: $\Delta R=50\,\mathrm{nm}$ (blue solid line), and  $\Delta R=20\,\mathrm{nm}$ (orange symbols), as a function of the laser intensity. The values of the intensity are multiplied by $20/50 = 0.4$ for the latter. (f) Propulsion speed $U$ at $T_{\infty} = 25^{\circ}$~C as a function of heating intensity $I$ for particles with cap thickness $\Delta R = 50$~nm in the linear regime $U \propto I \Delta R$ and radius $R= 1.37\,\mu\mathrm{m}$ ($\triangleleft$), $1.63\,\mu\mathrm{m}$ ($\triangledown$),  $2.14\,\mu\mathrm{m}$ ($\bigtriangleup$), $2.45\,\mu\mathrm{m}$ ($\times$), $3.14\,\mu\mathrm{m}$ ($\square$), $3.88\,\mu\mathrm{m}$ ($\circ$), $8.02\,\mu\mathrm{m}$ ($\ast$), and $11.78\,\mu\mathrm{m}$ ($\diamond$). The gray shaded area represents the experimental error of $U$.}
\label{fig:fig1}
\end{figure}

The active colloids in our experiments are spherical SiO$_2$ particles, with a radius that varies from $R = 1.37\,\mu\mathrm{m}$ to $R = 11.78\,\mu\mathrm{m}$, half-coated by a carbon cap of thickness $\Delta R = 50$~nm. Such an anisotropy allows one to define the particle orientation  $\mathrm{{\bf e}}$ as the unit vector pointing from the capped to the uncapped hemisphere, as illustrated in Figs.~\ref{fig:fig1}(a)-(b). The particles are suspended in a binary mixture of water and propylene glycol n-propyl ether (PnP), whose lower critical point is $T_c = 31.9^{\circ}$~C and 0.4 PnP mass fraction~\cite{Bauduin:2004}, and its viscosity at $25^{\circ}$  is $\eta_f = 0.004$~Pa~s. The phase diagram of this binary mixture is shown in Fig.~\ref{fig:fig1}(c). A dilute particle suspension is confined in a sample cell made of two glass slides and maintained at constant bath temperature, $T_{\infty}$, which can be controlled by a thermostat (accuracy of 0.02~K) and kept below the critical temperature, $T_{\infty} < T_c$, i.e. in the one-phase state.  In all our experiments, the aspect ratio between the separation $h$ between the two confining walls of the cell and the particle radius $R$ is fixed at $h/R =4$ in order to keep the same local hydrodynamics conditions. In such a confinement, the translational $D_t$ and rotational $D_r$ diffusion coefficients of the spherical colloids are smaller than the values $D_t^0 = k_B T_{\infty}/(6\pi\eta_f R)$ and $D_r^0 = k_B T_{\infty}/(8\pi \eta_f R^3)$ in the bulk: $D_t / D_t^0 \approx 0.4$ and $D_r / D_r^0 \approx 0.9 $, respectively. 

In order to induce self-propulsion, a uniform laser illumination ($\lambda = 532\,\mathrm{nm}$), whose intensity $I$ can be accurately adjusted, is perpendicularly applied onto the sample cell, as sketched in Fig.~\ref{fig:fig1}(a). Due to the high light absorption by the carbon cap compared to that of silica and the surrounding fluid, the temperature non-isotropically rises around the particle surface. Local demixing of the binary fluid around the particle does not occur unless the fluid temperature exceeds $T_c$, in which case self-propulsion can be induced by chemical potential gradients along the particle symmetry axis~\cite{Samin:2015}. 
Indeed, we observe that above an intensity of $I_0 \sim 0.5 \,\mu\mathrm{W}\,\mu\mathrm{m}^{-2}$, all the particles we investigate perform active Brownian motion~\cite{Buttinoni:2012,Samin:2015,Wuerger:2015}, where the translational and rotational dynamics takes places in two dimensions. In such a case, the 2D translational mean-square displacement~\cite{tenHagen:2011} has a diffusive contribution and a ballistic term due to self-propulsion at velocity $\mathbf{U}$,
\begin{equation}\label{eq:msd}
 	\langle |\mathrm{{\bf r}}(t) - \mathrm{{\bf r}}(0)|^2\rangle = 4D_t t + U^2 t^2, \,\,\, t \ll D_r^{-1},
\end{equation}
from which we extract the propulsion speed $U = \mathbf{U}\cdot \mathbf{e}$. (See Methods for further details regarding the experiment).

In Fig.~\ref{fig:fig1}(d), we plot as a solid line the typical dependence of the propulsion speed $U$ on the incident laser intensity $I$ for a particle of radius $R = 3.88\,\mu\mathrm{m}$ at bath temperature $T_{\infty} = 25^{\circ}$~C. Interestingly, in contrast to catalytic and thermophoretic colloidal microswimmers, where $U$ is a monotonically increasing function of the fuel concentration~\cite{Howse:2007} and the heating intensity~\cite{{Bregulla:2015}}, respectively, here the dependence on $I$ is strongly non-monotonic. For instance, only at sufficiently small $I$, $U$ increases linearly with the applied intensity. This linear self-propulsion mechanism was proposed  in~\cite{Buttinoni:2012} and has been recently used to experimentally investigate active motion in viscoelastic fluids~\cite{GomezSolano:2016} and in light gradients~\cite{Lozano:2016}. In such a case, the direction of the swimming velocity  is parallel to the particle orientation $\mathbf{e}$, $\mathbf{U} = |U|\mathbf{e}$, as sketched in Fig.~\ref{fig:fig1}(d), i.e., the particle moves away from the coated cap. This observation suggests that, for $I > I_0$, similar to thermophoretic active colloids~\cite{Jiang:2010,Bregulla:2015}, the swimming speed, $U$, linearly depends on the temperature gradient across the particle surface~\cite{Bickel:2013}, which is proportional to absorbed power $\sigma I$ divided by the geometrical factor $R^2$, where $\sigma$ is the absorption cross-section of the carbon cap~\cite{Bregulla:2015}. Since $\sigma$ scales with the volume of the cap $\propto R^2 \Delta R$, we have 
\begin{equation}
\label{eq:I}
	U \propto I\Delta R~,
\end{equation}
for a cap thickness $\Delta R \ll R$. Indeed, we checked that, at fixed $I$, by reducing the cap thickness from $\Delta R =50$~nm to $\Delta R =20$~nm, i.e. a factor 0.4, the propulsion speed is accordingly reduced, as shown in Fig.~\ref{fig:fig1}(e). In Fig.~\ref{fig:fig1}(f), we plot the dependence of the propulsion speed on the heating intensity for different particle radii ranging from $R = 1.37\,\mu\mathrm{m}$ to $R = 11.78\,\mu\mathrm{m}$ and constant $\Delta R = 50$~nm, where the same linear dependence $ U \propto I$ holds for all $R$, in agreement with the scaling~(\ref{eq:I}) for $U$. This is in stark contrast with the behavior of catalytic microswimmers, for which $U \propto R^{-1}$ at a given fuel concentration and cap thickness~\cite{Ebbens:2012}. 

Further increasing the heating intensity $I$ leads to strong deviations from the linear behaviour $U \propto I$ for all particle sizes at constant $\Delta R = 50$~nm. For instance, for $R = 3.88\,\mu\mathrm{m}$ the propulsion velocity starts to level off to $U \approx 1.2\,\mu\mathrm{m}\,\mathrm{s}^{-1}$ at $I \approx 12\,\mu\mathrm{W}\,\mu\mathrm{m}^{-2}$, as shown in Fig.~\ref{fig:fig1}(d). In addition, we find an abrupt decrease of the absolute value of the speed at a certain intensity threshold at which self-propulsion is suppressed to $U=0$. In the following, we denote this intensity as $I_{rev}$. Unexpectedly, at  $I > I_{rev}$, the propulsion speed becomes finite again but the swimming directionality reverses: the propulsion velocity is in this case anti-parallel to $\mathbf{e}$,  $\mathbf{U} = -|U|\mathbf{e}$, i.e., the particle moves toward the carbon cap, as schematized in Fig.~\ref{fig:fig1}(d). Higher values of $I$ above $I_{rev}$ give rise to a second non-monotonic propulsion behavior, where $|U|$ exhibits a global maximum while the particle subsequently comes to a halt as $I \rightarrow \infty$. For example, for a particle with radius $R = 3.88\,\mu\mathrm{m}$, the directional reversal occurs at $I_{rev} \approx 26\,\mu\mathrm{W}\,\mu\mathrm{m}^{-2}$ while the global maximum in $|U|$ is reached at $I \approx 30\,\mu\mathrm{W}\,\mu\mathrm{m}^{-2}$, at which $U \approx -5\,\mu\mathrm{m}\,\mathrm{s}^{-1}$. In this regime, the suppression of self-propulsion is more pronounced for bigger particles, for which cessation of directed-motion occurs at relatively small intensities, as observed for a  $R = 11.78\,\mu\mathrm{m}$ particle at $I \gtrsim 12 \,\mu\mathrm{W}\,\mu\mathrm{m}^{-2}$. Note that, unlike the linear self-propulsion at low laser intensities, the resulting swimming speed in this non-linear regime is strongly dependent on the particle size. For larger particles, the plateau at $I < I_{rev}$ becomes smaller, while at $I > I_{rev}$ the velocity reversal is shifted to lower values of $I$. 
In addition, we checked that the change in directionality is completely reversible: by decreasing the intensity from $I > I_{rev}$ back to  $I < I_{rev}$, the particle moves again with the cap at its rear relative to the swimming direction, where the response time of the swimming directionality to a change of $I$ is almost instantaneous for our temporal resolution. This reversibility is due to the large thermal diffusivity $\alpha_f \sim 10^{-7}\,\mathrm{m}^2\mathrm{s}^{-1}$ of the fluid, such that the temperature field responds to illumination intensity changes within a time scale $\lesssim R^2/\alpha_f \sim 10^{-5}\,\mathrm{s}$, thereby adjusting almost immediately to composition and velocity changes.

 \begin{figure}[ht]
	\includegraphics[width=\linewidth]{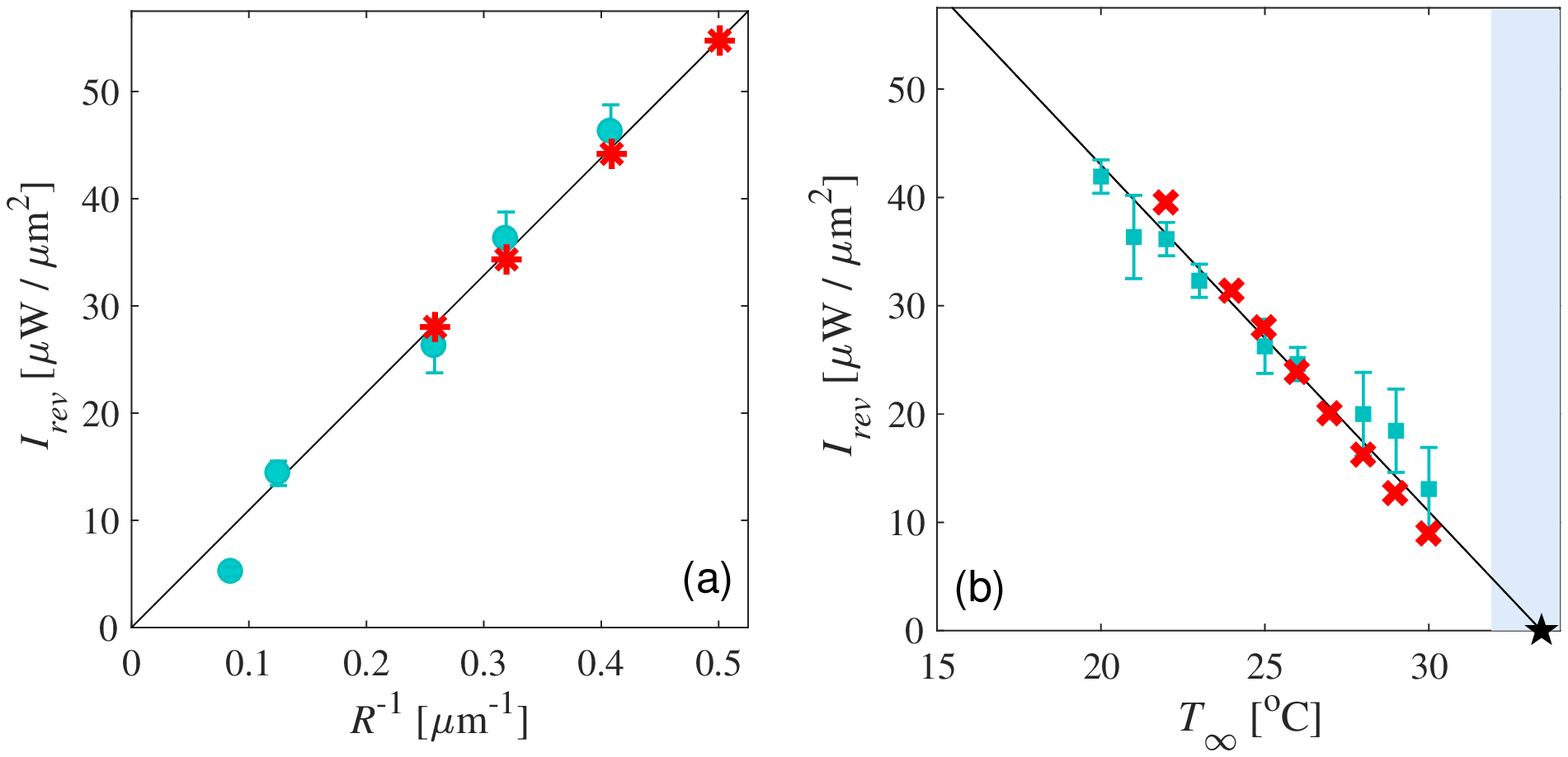}
	\caption{(a) Dependence of the minimum heating intensity $I_{rev}$ to induce directional reversal  as a function of the inverse particle radius $R^{-1}$ at $T_{\infty} = 25^{\circ}$~C, determined experimentally ($\circ$) and numerically ($\ast$). (b) Dependence of $I_{rev}$ as a function of the environment temperature $T_{\infty}$ for a $R = 3.88\,\mu\mathrm{m}$ particle obtained experimentally ($\square$) and numerically ($\mathbf{\times}$). The black star represents the minimum temperature $T_0 \approx 34^{\circ}$~C of the carbon cap at which the self-propulsion reverses its directionality.}
\label{fig:fig2}
\end{figure}

To characterize the transition between these two swimming regimes with distinct directionality, we investigate the role of the particle size and the bulk temperature $T_{\infty}$ of the binary liquid in the behaviour of the intensity threashold $I_{rev}$. Such a directional reversal at which $U=0$, must be related to a qualitative change in the shape of the demixed liquid around the colloid at a given temperature $T_0>T_c$ of the heated carbon cap regardless of the particle size. As a matter of fact, the maximum local temperature increase $\Delta T = T_{max} - T_{\infty}$ of the fluid around the particle is proportional to the absorbed power $\sigma I$ divided by the particle size $R$:  $\Delta T \propto \sigma I / (k_f R)$, where $k_f$ is the thermal conductivity of the fluid. Since the absorption cross section $\sigma$ is proportional to the volume of the cap, $R^2 \Delta R$, then for a fixed cap thickness $\Delta R$ the temperature increase scales as $\Delta T \propto R I$, which implies for $T_{max} = T_0$ that
\begin{equation}
\label{eq:Irev}
	I_{rev} \propto \frac{T_0 - T_{\infty}}{R}~.
\end{equation}
In agreement with (\ref{eq:Irev}), in Figs.~\ref{fig:fig2}(a) and \ref{fig:fig2}(b) we experimentally show that $I_{rev} \propto R^{-1}$ for particles of various radii at $T_{\infty} = 25^{\circ}$C, while $I_{rev} \propto T_0 - T_{\infty}$ for a particle of radius $R = 3.88\,\mu\mathrm{m}$ at different bath temperatures, respectively.

In order to better understand this remarkable swimming behaviour, we numerically investigate the local phase ordering of the PnP-H$_2$O mixture around the half-coated particle by means of  a non-isothermal diffuse-interface approach~\cite{anderson1998,Samin:2015}. 
Here, we denote by $\phi$ the volume fraction of PnP in the mixture ($0<\phi<1$). 
For a near-critical mixture the natural order parameter is the deviation $\varphi=\phi-\phi_c$, where $\phi_c$ is the critical 
composition of the mixture. The Ginzburg-Landau free energy for the  mixture is written as $F=\int {\rm d} \mathbf{r} f$, where the symmetric free 
energy density $f(\varphi,T)$ is:
\begin{equation}
\label{eq:f0}
\frac{V_0}{k_BT}f=2\frac{T_c-T}{T}\varphi^2+\frac{4}{3}
 \varphi^4+\frac{C}{2}|\nabla\varphi|^2~.
\end{equation}
Here, $k_B$ is the Boltzmann constant and $V_0 = a^3$ is the molecular volume, assumed equal for both mixture components. The first two terms in (\ref{eq:f0}) constitute the bulk free energy and give a lower critical solution temperature type phase diagram. The gradient term in (\ref{eq:f0}) accounts for the energetic cost of composition 
inhomogeneities, where $C=2a^2T_c/T$~\cite{safran}.

Using (\ref{eq:f0}) we can calculate the mixture chemical potential 
 $\mu=V_0(\delta F / \delta \varphi )$, which means:
\begin{equation}
\label{eq:mu}
\frac{\mu}{k_BT}=4\frac{T_c-T}{T}\varphi+\frac{16}{3}
 \varphi^3-C\nabla^2\varphi~.
\end{equation}
In equilibrium, $\mu$ is homogeneous throughout the system, but during phase 
ordering chemical potential gradients develop leading to composition currents. 
The composition kinetics is described by the continuity equation for 
$\varphi \in [-1/2,1/2]$, known as the Cahn-Hilliard equation:
\begin{equation}
\label{eq:ch}
 \frac{\partial \varphi}{\partial t} = -\nabla \cdot \mathbf{j_\varphi}=- \nabla 
\cdot \left( \varphi \mathbf{v}
-D \nabla \frac{\mu}{k_BT} \right) ~,
\end{equation}
where $D$ is the mixture inter-diffusion constant and $\mathbf{v}$ is the
fluid velocity. The composition current ${\mathrm{\bf{j_\varphi}}}$ in (\ref{eq:ch}) 
is composed of a convective term $ \varphi \mathbf{v}$ and diffusive term $\propto \nabla \mu/T$.
We investigate the motion of the active colloid during the enhanced diffusion trajectory. A steady
state of the composition is achieved quickly during this motion, since the natural
time-scale for the phase separation kinetics around the colloid as described by Eq.~(\ref{eq:ch}), 
$R^2 / D \sim 0.1 - 1$~s, is much shorter than the particle rotational diffusion time
$D_r^{-1}\sim 10^2 - 10^4$~s. This is also confirmed by our numerical calculations.

Because the Reynolds number is ${\rm Re} =\rho_f R U/\eta_f \ll 1$, where the fluid density, $\rho_f$, and viscosity, 
$\eta_f$, are assumed constant, the dynamics of the liquid around the microswimmer is governed by the 
Stokes equations for an incompressible fluid:
\begin{eqnarray}
\label{eq:nc}
\nabla \cdot \mathbf{v} & = &0~,\\
\label{eq:ns}
 \eta\nabla^2 
\mathbf{v} & = & \nabla p+\frac{\varphi}{V_0} \nabla\mu~,
\end{eqnarray}
where the pressure $p$ follows from the incompressibility condition (\ref{eq:nc}). 
The last term in (\ref{eq:ns}) is the capillary body force due to chemical 
potential gradients, and is well known in critical dynamics, 
where (\ref{eq:ch})-(\ref{eq:ns}) are also known as ``model H'' 
\cite{hohenberg1977}.  Notice that, when the fluid is incompressible, we  can rewrite the body forces on the right hand side of (\ref{eq:ns}) as $-\nabla 
 p'- \mu \nabla \varphi/V_0$, where $p'$ is an effective pressure, which is easier to treat numerically.

Heat diffuses within the fluid and the solid much faster than the mixture 
components inter-diffuse. Hence, the temperature field in and around the particle 
adjusts immediately on the time scale for the composition dynamics. 
Moreover, the advection of heat in the fluid can be neglected since 
the thermal P\'{e}clet number, $\mathrm{Pe}_T=U R/\alpha_f\ll 1$. Thus, the temperature 
field simply follows the particle as it translates, adjusting immediately 
to composition and velocity changes. This means that the heat equation in both the solid 
and liquid reduces to the Laplace equation 
\begin{equation}
\label{eq:lap}
\nabla^2 T=0~, 
\end{equation}
with a constant heat flux $q_0$ at the poorly conducting carbon cap. For more computational details, all of the boundary conditions for the solution of Eqs.~(\ref{eq:mu})-(\ref{eq:lap}), and obtaining the particle velocity $U$ from the numerical solution, see the Methods section. 

\begin{figure}[ht]
	\includegraphics[width=\linewidth]{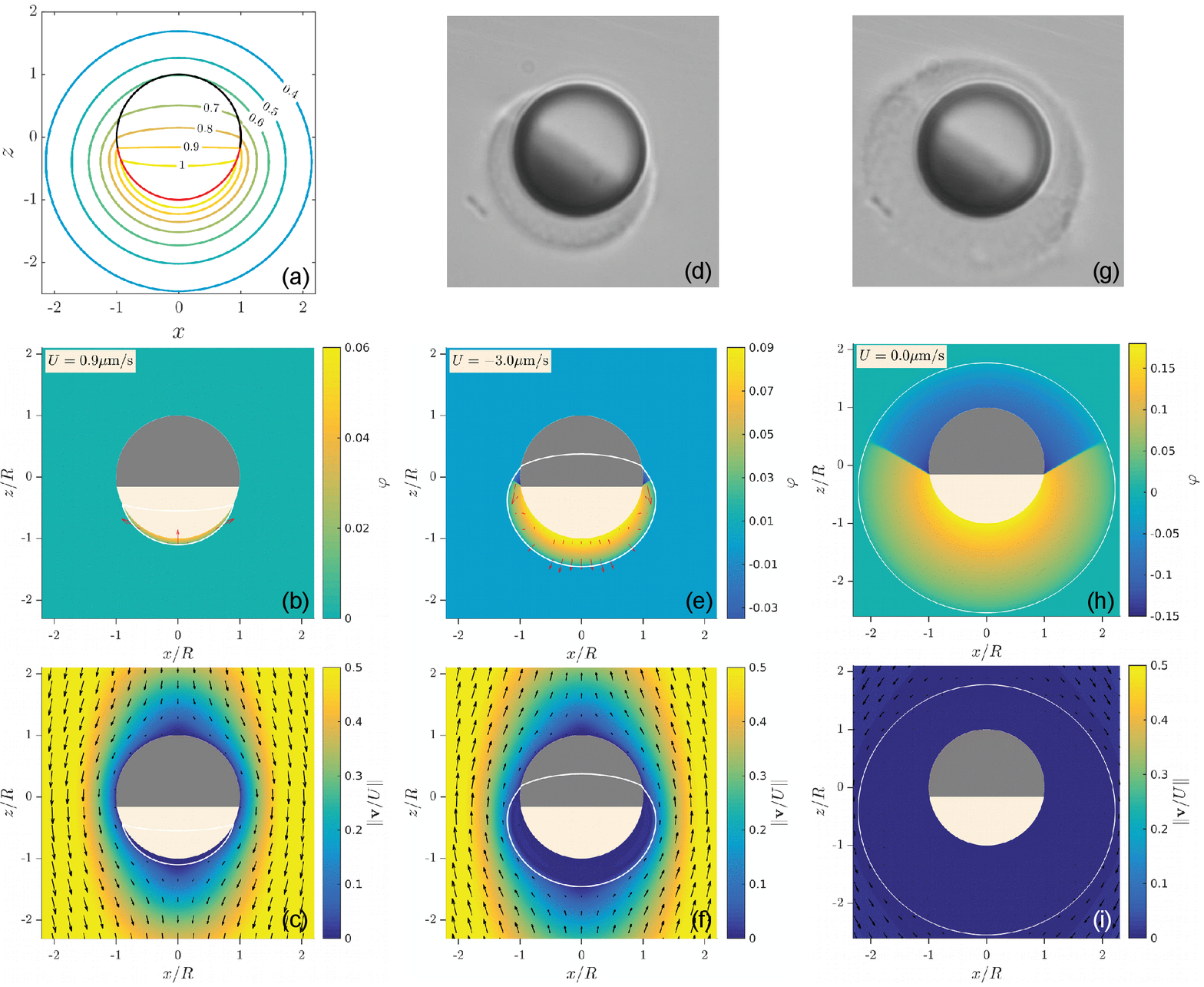}
	\caption{(a) Reduced temperature contours in the $(xOz)$ plane around a swimmer 
with a constant heat flux boundary condition at the carbon cap (red line) and a solid-fluid heat conductivity contrast $K \approx 
2.5$. (b) Steady state composition $\varphi$ and (c) normalized velocity magnitude $||\mathbf{v}/U||$ around a particle immersed in a mixture with a temperature $T_{\infty}=25^{\circ}$~C and illuminated such that $I<I_{rev}$. 
The white line is the critical temperature contour, and the swimming velocity $U$ appears in the top left corner. In (b) the arrows indicate the body-force and in (c) the arrows are velocity vectors. (d) Snapshot of an active particle in the binary mixture at $T_{\infty} = 25^{\circ}$~C at $I>I_{rev}$.  (e) The corresponding steady state composition $\varphi$ and (f) velocity magnitude $||\mathbf{v}/U||$ obtained numerically. (g) Snapshot of an active particle in the binary mixture at $T_{\infty} = 25^{\circ}$~C at $I\gg I_{rev}$. (h) The corresponding steady state composition $\varphi$ and (i) velocity magnitude $||\mathbf{v}/U||$ obtained numerically.}
\label{fig:fig3}
\end{figure}

We first verify that the model given by Eqs.~(\ref{eq:mu})-(\ref{eq:lap}) reproduces the main experimentally observed features of the heating intensity $I_{rev}$, which determines the threshold for the reversal of the swimming direction. By numerically finding the smallest non-zero heat flux $q_0$ across the carbon cap needed to obtain a zero steady-state propulsion speed, we determine $q_{rev}$ for different values of the particle radius $R$ and the bath temperature of the fluid $T_{\infty}$. Our numerical results show that, $q_{rev} \propto T_0 - T_{\infty}$, where $T_0 > T_c$, and $q_{rev} \propto R^{-1}$, in excellent accordance with the relation (\ref{eq:Irev}). Due to the heat loss of the incident illumination through the sample cell, $q_{rev} \lneqq I_{rev}$. In fact, in Figs.~\ref{fig:fig2}(a) and (b) we show that the numerical and the experimental results have a very good agreement if we set $q_{rev}=\kappa I_{rev}$, with $\kappa \approx 0.13$, for both dependences on $R$ and $T_{\infty}$, respectively.

The reduced temperature field around the particle, obtained by numerically solving (\ref{eq:lap}) with the proper boundary conditions, is shown in Fig.~\ref{fig:fig3}(a). The temperature is maximal at the heated hemisphere and decays in the radial direction and along the colloid contour. When $I$ is large enough, the temperature at the colloid surface will exceed the critical temperature. In this case, at some distance from the colloid surface, a temperature contour of the critical temperature $T_c$ will enclose a region for which $T>T_c$. Demixing of the fluid occurs within this region, such that the shape of the temperature contours determines the resulting steady-state swimming \cite{Samin:2015}.  
For instance, in Fig.~\ref{fig:fig3}(b) we show the steady-state profile of the composition $\varphi$ for a heating laser intensity $I < I_{rev}$.
Within the $T_c$ isotherm (white curve), a single PnP-rich ($\varphi > 0$) droplet is nucleated at the hydrophobic carbon cap, partially covering it. 

In Fig.~\ref{fig:fig3}(c), we plot the corresponding flow field of the mixture in the reference frame of the particle. 
Within the demixed region, the fluid velocity is very small, revealing that the droplet moves together with the particle as it self-propels in the laboratory frame. Therefore, unlike catalytic and thermophoretic Janus colloidal microswimmers, here, a slip-velocity on the particle surface is not responsible for the resulting self-propulsion. Instead, self-propulsion is generated by body forces, $\propto \mu \nabla \varphi$, which are localized at the droplet edges where composition gradients are large, see the arrows is Fig.~\ref{fig:fig3}(b). Since the droplet extends far from the colloid surface, at a distance that is of the order of $R$, the thin-layer approximation cannot be applied, and the common picture of a surface slip velocity is not suitable. Rather, the pressure gradient generated by the anisotropic body forces in the demixed droplet is transmitted to the colloid surface, exerting a force perpendicular to the surface, thereby leading to directed motion~\cite{Samin:2015}. The fluid's velocity field outside and far from the droplet, shown in Fig.~\ref{fig:fig3}(c), points in the negative $z$ direction, which corresponds to a swimming velocity in the direction opposite to the carbon cap, in agreement with our experimental observations.

We now consider self-propulsion for $I > I_{rev}$. In Fig.~\ref{fig:fig3}(d) we show a snapshot of this experimental situation where the propulsion velocity exhibits the opposite directionality, i.e. the particle moves with the cap at the front. In such a case, we can clearly observe a large asymmetric droplet around the particle surface. The droplet can be easily visualized due to the difference of refractive index between PnP an the homogeneous binary mixture. The corresponding steady-state profile of $\varphi$ computed from the numerical solution exhibits a similar shape. Interestingly, in addition to the PnP-rich droplet that nucleates around the capped site, a second water-rich droplet is nucleated at the hydrophilic silica hemisphere, see Fig.~\ref{fig:fig3}(e). We attribute the reversal of the swimming direction to this qualitative change, which occurs when the temperature at the \emph{non}-heated cap increases above $T_c$. Our calculations reveal that the nucleation of the second droplet leads to an oppositely directed body force near the new liquid-liquid interface, and thus to a change in the self-propulsion direction, see Figs.~\ref{fig:fig3}(e)-(f). 
Although the water droplet in Fig.~\ref{fig:fig3}(e) is much smaller than
the PnP droplet, the body forces due its nucleation are much closer to the particle
surface, see the vectors in Fig.~\ref{fig:fig3}(e), and thus are able to reverse the self-propulsion
direction. Furthermore, at sufficiently strong heating intensity above $I_{rev}$, the demixed region completely encloses both hemispheres of the particle, as we observe in Fig.~\ref{fig:fig3}(g), which also evidences the nucleation of two distinct droplets clearly. When $I_{rev}$ further increases, $|U|$ begins to eventually decrease. When the droplet thickness is comparable to, or larger than $R$, its shape becomes essentially radially symmetric with respect to the particle center, and self-propulsion becomes negligible. This is illustrated in Fig.~\ref{fig:fig3}(g)-(i), where we compare the experiments with the numerical solutions of the steady state $\varphi$ and velocity map, confirming that in this case $|U|\approx 0$.

\begin{figure}[ht]
	\centering
	\includegraphics[width=0.85\linewidth]{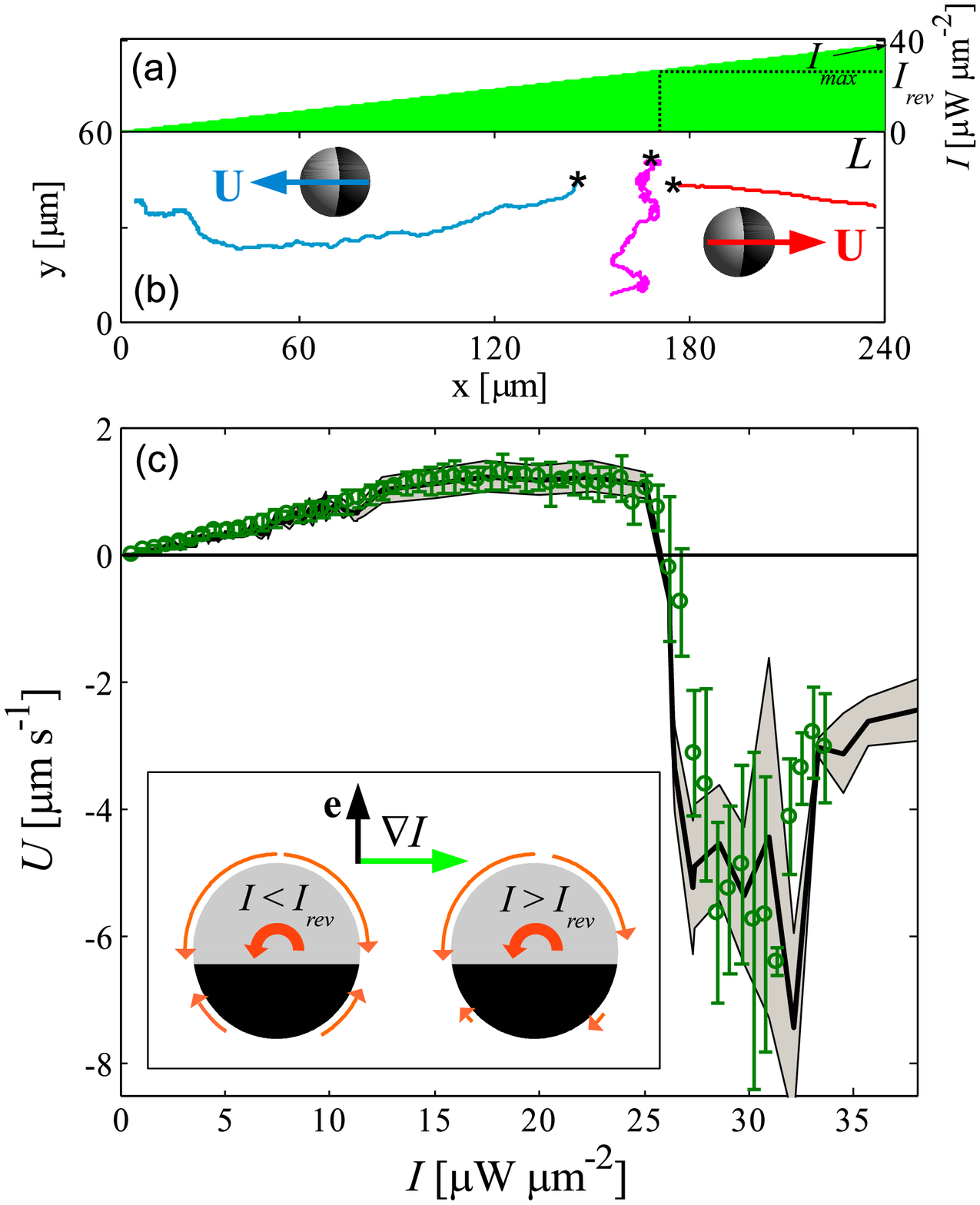}
	\caption{(a) Light-intensity profile along the $x$-direction, linearly increasing from 0 to $I_{max} = 38\,\mu\mathrm{W}\,\mu\mathrm{m}^{-2}$ over a distance $L = 240\,\mu\mathrm{m}$. (b) Examples of trajectories of an active colloid (radius $R = 3.88 \, \mu\mathrm{m}$, bath temperature $T_{\infty} = 25^{\circ}$~C) moving at different locations in such a light field. The stars indicate the starting position $(x_0,y_0)$ of the particle.  (c) Dependence of the propulsion speed $U(x,y)$ as a function the local heating intensity $I(x,y)$ with uniform gradient $\nabla I$ ($\circ$), compared to that measured in presence of uniform illumination (solid line). Inset: schematic representation of the mechanism which leads to negative ($I < I_{rev}$) and positive ($I > I_{rev}$) phototaxis in a linear intensity gradient $\nabla I$. A particle, whose orientation $\mathbf{e}$ (vertical black arrow) is perpendicular to the gradient (green horizontal arrow), experiences a non-zero torque, where the corresponding integrated torque densities on each quadrant are represented by the curved arrows.
The direction and length of the arrows are based on the numerical calculations, whose lengths are proportional to the relative magnitud of the different torque densities, normalized such that the largest value is 1. In both cases, the net torque tends to rotate the particle counter-clockwise, thus leading to an antiparallel alignment of $\mathbf{e}$ with respect to $\nabla I$, in which case the total torque vanishes.}
\label{fig:fig5}
\end{figure} 

Finally, we experimentally investigate the 2D motion of a half-coated particle in an area $L\times L$ in the presence of a linear intensity profile, depicted in Fig.~\ref{fig:fig5}(a): $I(x,y) = I_{max}x/L$, where $I_{max} = I(L,y) > I_{rev}$,  which corresponds to a uniform intensity gradient $\nabla I = (I_{max}/L,0)$.
In Fig.~\ref{fig:fig5}(b) we show some exemplary trajectories of an active colloid (radius $R = 3.88\,\mu\mathrm{m}$) moving through the binary mixture kept at $T_{\infty} = 25^{\circ}$~C, where $I_{max} = 38\,\mu\mathrm{W}\,\mu\mathrm{m}^{-2}$, $L = 240\, \mu\mathrm{m}$ and $I_{rev} = 26 \,\mu\mathrm{W}\,\mu\mathrm{m}^{-2}$. Under such conditions, the orientation $\mathbf{e}$ clearly reveals the existence of positive and negative phototaxis, i.e. the particle is able to sense the gradient direction and to move either toward or away from it. This phototactic behavior is characterized by a large reduction of the randomness of the orientational dynamics due to rotational diffusion, where $\mathbf{e}$ becomes strongly oriented along the gradient~\cite{Lozano:2016,Bennett:2015}. 

Strikingly, we find that three kinds of phototactic responses to the same uniform $\nabla I$ occur depending on the initial particle position $(x_0, y_0)$ on the light field. If the initial position is such that the local intensity is $I(x_0,y_0) <  I_{rev}$, the particle exhibits negative phototaxis, i.e. it self-propels in the direction opposite to $\nabla I$ with the cap at its rear. On the other hand, when the particle initially experiences an intensity $I(x_0,y_0)>I_{rev}$, it displays positive phototaxis, moving toward the gradient with the cap at the front. In addition, within a narrow region where the local intensity is $I(x,y) \approx I_{rev}$, the orientational response is  not clearly defined and the particle can even undergo large displacements along the $y$-direction perpendicular to $\nabla I$. The particle eventually escapes this unstable region and moves far away from it with one of the two well-defined tactic behaviors. We note that, for a uniform gradient, the coexistence of these three distinct orientational responses is very uncommon for self-diffusiophoretic microswimmers, where a single tactic response, either positive, negative or none, occurs in absence of any elaborate external steering or feedback mechanism~\cite{Saha:2014}. 
We attribute the observed behavior in our experiments to the non-monotonic dependence of the velocity on the heating intensity, where $I_{rev}$ separates two distinct regions on the light field, where the same orientational response leads to propulsion in opposite directions.

To understand this behavior, we performed calculations of a simplified 2D system of a heated disk in a linear temperature gradient, since our 3D axisymmetric calculations do not allow for a net torque on the particle. These calculations revealed that in both swimming regimes, as expected, the temperature gradient results in the nucleation of asymmetric droplet(s) at the disk surface. Integration of the resulting torque distribution on the disk surface leads to an antiparallel reorientation in both cases, as observed in experiments. The behavior is caused by the fact that the hemisphere which is down the gradient (with respect to $\mathbf{e}$) experiences a larger torque, see the schematic arrows at the particle surface in the inset of Fig.~\ref{fig:fig5}(c). Due to the dependence of the swimming directionality on the local intensity $I$, the particle then moves with the cap at the rear to regions of low intensity for $I(x,y) <  I_{rev}$, as observed in~\cite{Lozano:2016} for very low heating intensities. For $I(x,y) >  I_{rev}$, however, the particle self-propels toward regions of high intensity with the cap at the front.

In fact, we verify that, once the particle re-orients and reaches a stable angular configuration, its resulting propulsion velocity $\mathbf{U}(x,y)$, which in a gradient is position-dependent, is only determined by the local intensity $I(x,y)$. This is demonstrated in Fig.~\ref{fig:fig5}(c), where we show an excellent quantitative agreement between the results for $U$ obtained under uniform illumination, and those in presence of the linear light-intensity profile. We stress that such a unique tactic behavior is due to the nature of self-propulsion by the demixing, whereby the complex hydrodynamics of the continuously demixed droplet determines the response to the gradient, and is currently absent in the case of self-phoretic microswimmers with an effective surface velocity~\cite{Saha:2014}.

\section*{Discussion}

Microorganisms have evolved to developed many navigation strategies based on internal molecular processes that allow them to adapt their motility to specific fluid environments.
Among these strategies, {\it run-and-reverse} and {\it run-and-reverse-flick} motion~\cite{Stocker:2011}, where the swimming direction can be completely reversed, lead to large enhancements of spreading and the formation of complex spatiotemporal patterns, 
that would be otherwise absent in the case of e.g. simple of {\it run-and-tumble} of {\emph{Escherichia coli}}. 
In a similar way, reversible {\it in-situ} tuning of the velocity of synthetic microswimmers by means of simple physical rules is an appealing property for the design of autonomous  microrobotic devices in e.g. biomedical applications, where a complete reversal of the motile directionality is required to differentiate, exploit, or overcome particular environmental conditions. 

In this article, we experimentally and numerically demonstrate that such a reversible tuning of the propulsion velocity is feasible for active half-coated colloids
suspended in critical binary mixtures. Although a number of mechanisms have been recently proposed in order to explain the swimming directionality in binary mixtures, e.g. diffusiophoretic and surface charge effects~\cite{Wuerger:2015}, our findings clearly show that those factors are not dominant in our system. Indeed, in our experiments, finely tunable laser-heating gives rise to asymmetric chemical potential gradients in the mixture around the particle, which in turn lead to directed motion. Such chemical potential gradients are generated inside a droplet nucleated around the colloid within the isotherm $T = T_c$, and result in body forces far from the particle surface, with no counterpart in other phoretic colloidal microswimmers, for which self-propulsion is achieved by slip flows at the particle surface. 

Our results reveal a non-monotonic behavior of the propulsion speed as a function of the heating intensity, which is attributed to the difference in the wetting of the two hemispheres, and which depends on the particle size, the cap thickness and the surrounding bulk temperature. Therefore, such parameters, in addition to the heat conductivity contrast between the solid and the liquid, can be smartly exploited to experimentally control the motility of colloidal microswimmers in more complex environments~\cite{GomezSolano:2015}.  
In addition, unlike other self-phoretic colloidal microswimmers, the strongly non-monotonic behavior of the propulsion speed along with its directionality reversal enable the possibility to realize, in a rather simple fashion, negative and positive phototaxis in a uniform light gradient. The key factor in the directional reversal of the particles is the fact that although only the capped hemisphere is heated, both
hemispheres have different wetting properties~\cite{Araki:2015,Dattani:2017}, and contribute to the self-propulsion in an opposite manner. Future work should address the exact
role of the wetability contrast, in particular, whether the reversal intensity depends
on this quantity. The presented mechanism could provide a design principle for other microswimmers, for instance in catalytic microswimmers, in the case where both hemispheres are made active, but catalyse two different reactions with different fuel molecules,
or the same reaction, e.g. with Michaelis-Menten kinetics~\cite{Howse:2007}, but with distinct surface reaction rates.
In such cases, external control over the fuel concentrations could possibly lead to a directional reversal. 
In summary, we demonstrated the complex swimming behavior of a synthetic self-propelled colloid, which allows to tailor its response the local environment. Our proof-of-concept experiments are a significant step forward for the design of the next generation of artificial microswimmers.

\section*{Methods}

\begin{figure}[ht]
\centering
\includegraphics[width=0.85\linewidth]{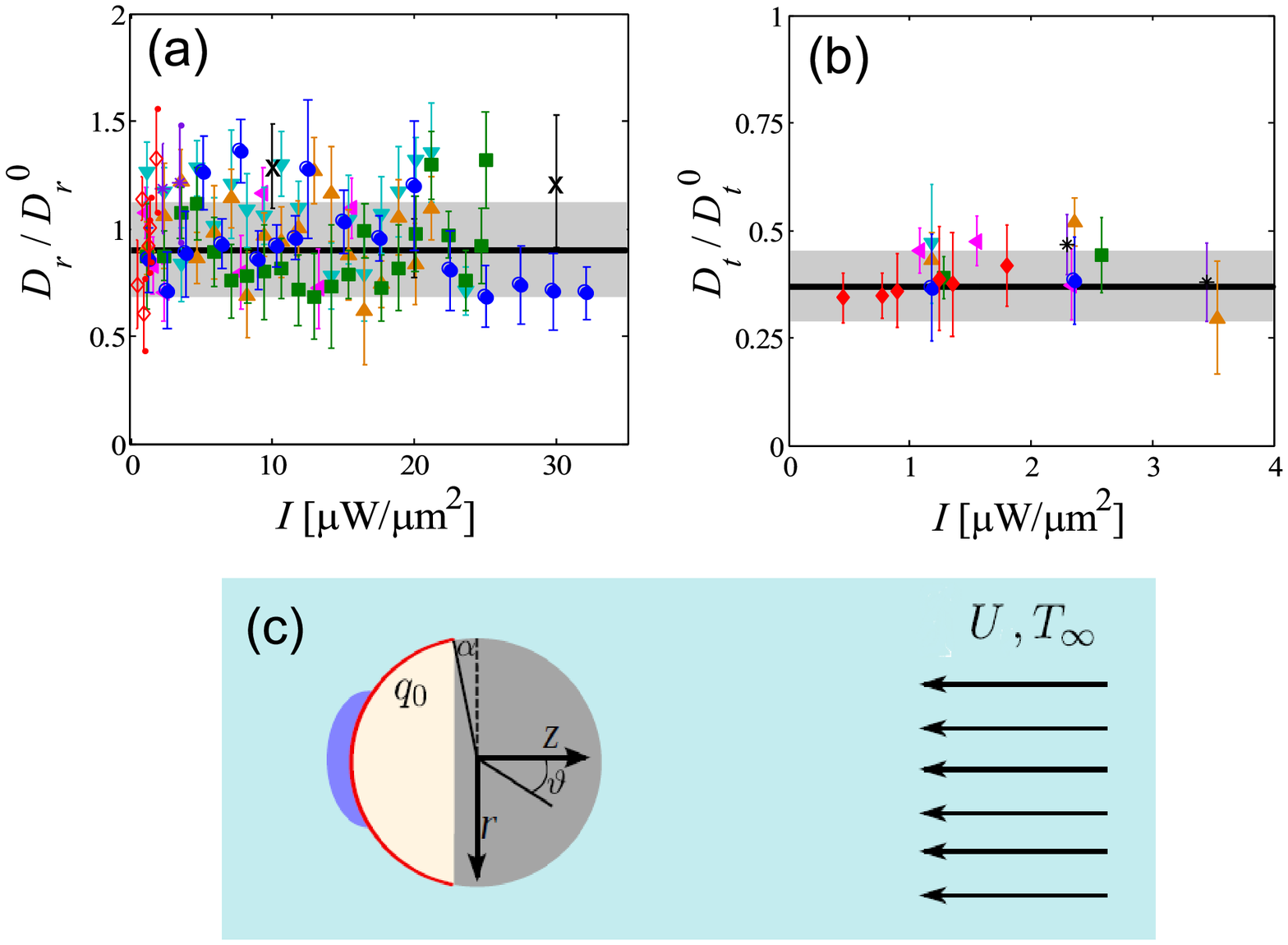}
\caption{
(a) Rotational and (b) translational diffusion coefficients of self-propelled colloids of different radii: $R= 1.37\,\mu\mathrm{m}$ ($\triangleleft$), $1.63\,\mu\mathrm{m}$ ($\triangledown$),  $2.14\,\mu\mathrm{m}$ ($\bigtriangleup$), $2.45\,\mu\mathrm{m}$ ($\times$), $3.14\,\mu\mathrm{m}$ ($\square$), $3.88\,\mu\mathrm{m}$ ($\circ$), $8.02\,\mu\mathrm{m}$ ($\ast$), and $11.78\,\mu\mathrm{m}$ ($\diamond$), normalized by the corresponding Stokes-Einstein values in the bulk, and measured at different illumination intensities. The solid lines and the shaded areas represent the mean and the error, respectively, of $D_r/D_r^0$ and $D_t/D_t^0$ measured in absence of activity and averaged over all particles sizes. 
(c) Schematic illustration of a Janus particle immersed in
a near-critical mixture with an ambient temperature $T_{\infty}$. Illumination of a 
thin carbon cap (red line) leads to a constant heat flux $q_0$ across the carbon cap. The ensuing demixing of a droplet near 
the cap propels the particle, such that the fluid velocity far from the particle 
is $U$ in a frame-of-reference co-moving with the particle.}
\label{fig:supp}
\end{figure}

\subsection*{Experimental description}

Synthetic half-coated colloidal microswimmers were produced from spherical silica particles (radius $R= 1.37\,\mu\mathrm{m}$, $1.63\,\mu\mathrm{m}$,  $2.14\,\mu\mathrm{m}$, $2.45\,\mu\mathrm{m}$, $3.14\,\mu\mathrm{m}$, $3.88\,\mu\mathrm{m}$, $8.02\,\mu\mathrm{m}$, and $11.78\,\mu\mathrm{m}$) by carbon sputtering.  The thickness of the cap was set to $\Delta R = 50$~nm in most of our experiments, but $\Delta R = 20$~nm caps were also created for specific purposes. 
The resulting two-faced particles were suspended in a binary mixture of H$_2$O and propylene glycol n-propyl ether (PnP) at the critical mass composition (60\% H$_2$O and 40\% PnP), whose temperature $T_{\infty}$ was adjusted between 20 and $30^{\circ}$~C by means of a thermostat. Images of the particles were acquired using video microscopy with a frame rate 25~fps  and submicrometric spatial resolution. 
From these images, the 2D positions $\mathbf{r} = (x,y)$ and the projections of the orientations $\mathbf{e} = (\cos \theta, \sin \theta)$ on the $x-y$ plane were obtained using the MATLAB image analysis software.

Because of the geometrical confinement vertically created by the two cell walls described in the main text, both the orientational and the translational dynamics of the self-propelled colloids are constrained in 2D, similar to the orientational quenching observed in other types of active particles~\cite{Das:2015}. Indeed,  we find that, while passive particles can perform 3D rotations, 2D rotational dynamics occur upon inducing self-propulsion. Consequently, the coordinates $x, y$ and $\theta$ are sufficient to describe the resulting active Brownian motion. In such a case, we verify that the dynamics of $\theta$ is purely diffusive with an angular mean-square displacement given by
\begin{equation}\label{eq:msdang}
 	\langle |\theta(t) - \theta(0)|^2\rangle = 2D_r t,
\end{equation}
regardless of the heating intensity $I$. 
In Fig.~\ref{fig:supp}(a) we plot the rotational diffusion coefficient $D_r$, obtained by fitting the experimental data to Eq.~(\ref{eq:msdang}), normalized by the bulk value $D_r^0 = \frac{k_B T_{\infty}}{8\pi \eta_f R^3}$ given by the Stokes-Einstein relation, as a function of $I$ for particles of different $R$. For comparison, in Fig.~\ref{fig:supp}(a) we represent $D_r / D_r^0$ measured for passive particles ($I=0$) as a solid line.
We observe that $D_r / D_r^0$ remains constant for all the illumination intensities and for all the particle sizes, which implies that the local laser heating does not significantly change the viscosity of the surrounding fluid. Due to the hydrodynamic interactions with the confining solid walls, the experimental values of $D_r$ are slightly smaller than $D_r^0$: $D_r / D_r^0 = 0.90 \pm 0.22$. In addition, we also compute the translational mean-square displacement $\langle |\mathrm{{\bf r}}(t) - \mathrm{{\bf r}}(0)|^2\rangle$ in order to obtain $D_t$ and the propulsion speed $U$ by fitting the experimental data to Eq.~(\ref{eq:msd})  under the condition $t \ll D_r^{-1}$,
In Fig.~\ref{fig:supp}(b) we show the resulting values of $D_t$, normalized by $D_t^0 = \frac{k_B T_{\infty}}{6\pi \eta_f R}$, for various particle sizes and laser intensities. In this case, we also find no significant dependence of $D_t$ on $I$, and that the presence of the confining walls lead to a translational friction much higher than that in the bulk: $D_t / D_t^0 \approx 0.37 \pm 0.08$.

\subsection*{Computational details}

For the numerical computations, it is convenient to non-dimensionalize the governing equations. 
We use the characteristic scales $R$ for length, $U$ for velocity, $R^2/D$ 
for time and $\eta_f U /R$ for pressure. We also introduce the scaled temperature $\Theta$ as $\Theta(T)=k_f(T-T_{\infty})/(q_0R)$,
where $q_0 \propto I$ is the heat flux across the carbon cap, induced by a uniform laser illumination.
In dimensionless form the governing equations, including the full heat equation, read:
\begin{eqnarray}
\label{eq:ch_dl}
  \frac{\partial \varphi}{\partial \tilde{t}} + \mathrm{Pe}_{\varphi} \tilde{\nabla} \cdot \left( \varphi 
\tilde{\mathbf{v}} 
\right)&=&\tilde{\nabla}^2 \mu~, \\
\label{eq:nc_dl}
\tilde{\nabla} \cdot \tilde{\mathbf{v}} &=&0, \\
\label{eq:ns_dl}
 \tilde{\nabla}^2 
\tilde{\mathbf{v}}- \tilde{\nabla} \tilde{p}  &=&  
 \frac{1}{\mathrm{CaC_h}}\varphi \tilde{\nabla} \mu, \\
 \frac{1}{\mathrm{Le}}\frac{\partial \Theta}{\partial \tilde{t}}+\mathrm{Pe}_T \tilde{\nabla} \cdot 
\left( \Theta \tilde{\mathbf{v}} \right) 
&=&
 \tilde{\nabla}^2 \Theta,
\label{eq:heat_dl}
\end{eqnarray}
Here, $\mathrm{Pe}_{\varphi}=U R/D$ is 
the composition P\'{e}clet number measuring the relative magnitude of 
advection to diffusion, while $\mathrm{Pe}_T=U R/\alpha_f$ is 
the thermal P\'{e}clet number, where $\alpha_f=k_f/(\rho_f C_f)$ is the 
fluid's thermal diffusivity, and $C_f$ is the fluid heat capacity. The typical radius of the microswimmers is 
$O(\mu$m$)$ and their swimming speed is $O(\mu$m/s$)$. This leads to $\mathrm{Pe}_T \ll 
1$, such that the Lewis number, $\mathrm{Le}=\mathrm{Pe}_{\varphi}/\mathrm{Pe}_T=\alpha_f/D$ is $\gg 
1$. Hence, we can safely neglect both the advection and 
the time dependence in (\ref{eq:heat_dl}). The same argument can be used for the solid phase, for which the solid thermal diffusivity 
$\alpha_s$ is of the order of $\alpha_f$. Therefore, the heat equation in both the solid 
and liquid reduces to the Laplace equation: $\tilde{\nabla}^2\Theta=0~$.
In (\ref{eq:ns_dl}),  $\mathrm{C_h}=a/R$ is the Cahn number and $\mathrm{Ca}=a^2\eta_f U/(k_BT)$ is the 
capillary number, measuring the relative magnitude of viscous and surface 
tension forces. $\mathrm{Ca} \ll 1$ for the typical velocities of the microswimmers, 
meaning that surface tension effects dominate the steady-state configuration. 

We consider a spherical colloidal particle with a radius $R$ and a thermal 
conductivity $k_s$ immersed in a homogeneous near-critical binary mixture 
having a temperature $T_{\infty}$ and thermal conductivity $k_f$. One side of 
the particles is coated with a thin layer of light-adsorbing carbon, having a 
thickness $\Delta R$, of the order of 50 nm, and a thermal conductivity $k_c$. For a 
very thin cap \cite{Bickel:2013} where $k_c/k_s, k_c/k_f < R/\Delta R$ both hold, we can 
neglect the cap thermal conductivity. Therefore, we assume that, when the 
particle is illuminated, there is a constant heat flux $q_0$ 
across the heated cap. Local demixing of the mixture occurs adjacent to the cap 
when its temperature crosses the critical temperature $T_c$ into the 
coexistence region of the mixture phase diagram of Fig.~\ref{fig:fig1}(c). The temperature 
profile is asymmetric with respect to the particle midplane leading to the 
demixing of a non-spherical droplet near the cap at steady-state, see the 
schematic illustration in Fig.~\ref{fig:supp}(c). Chemical-potential gradients within 
this droplet exert a non-isotropic force on the particle at the 
particle-droplet contact area, thus propelling it. At steady-state, the net 
force on the self-propelling particle $\mathbf{F}$ vanishes, and it attains a terminal swimming velocity 
$\mathbf{U}$. The problem's cylindrical symmetry, as illustrated in Fig.~\ref{fig:supp}(c), 
means that no net torque acts on the colloid. In 
a frame-of-reference co-moving with the particle, the fluid velocity far from 
the particle is axial: $\mathbf{U}=U \hat{\mathbf{z}}$.
In the illustration, the particle, placed at the origin of a cylindrical 
coordinate system, translates in the positive $z$ direction, with the cap at 
its rear. This is not the case in general.

In practice, the particles are not perfectly half-capped since the carbon sputtering method leads to slightly less than 50\% coverage,
Therefore, without loss of generality we assume 42\% coverage,
defined via the angle $\alpha=0.05\pi$, see the illustration in Fig.~\ref{fig:supp}(c). 
We verify that our results are qualitatively similar also for $\alpha<0$, i.e. for more than 50\% coverage, as
long as $|\alpha|$ remains small.
For the thermal conductivity of the silica colloid we used $k_s=1.38$ W/(m K).
The maximal temperature difference within the system is rather small, of the order of 1-10~K. We therefore make the approximation that most of the fluid's physical properties are independent of temperature. 
Even though $T_c$ lies within our  temperature window, this assumption is justified except for the inter-diffusion constant $D$, which vanishes as a power law close to $T_c$: $D=k_BT/(6\pi\eta_f\xi)$, where the bulk correlation 
length $\xi$ in our mean-field theory follows the scaling $\xi\propto(\left|T-T_c\right|/T)^{-\frac{1}{2}}$~\cite{Kawasaki:1970}. In the temperature window we examine, $D$ is of the order of 
$10^{-11}-10^{-12}$ m$^2$/s. For the other properties of the mixture we used in our calculations $T_c=31.9^{\circ}$~C, $\eta_f=4$ mP s, $k_f=0.56$ W/(m K), and a molecular size $a=3.7$ \AA.

Henceforth, the $\sim$ sign denoting reduced quantities will be omitted.
Azimuthal symmetry allows us to solve the problem of translational motion using a 2D axisymmetric 
cylindrical system $(r,z)$. The colloid is placed at the 
origin of a rectangular computational domain of length $2l=1000$ in $z$ 
($|z| \le l$) and width $l$ in $r$ ($0 \le r \le l$). The large domain size is required because the calculation of 
stresses on boundaries at zero Reynolds number is sensitive to the outlet and 
inlet flow boundaries.

For the symmetry axis at $r=0$ and the far away mixture at $r=l$, 
there are no fluxes normal to the 
boundaries. Therefore, the boundary condition (BC) for the 
composition is $\mathbf{n} \cdot \nabla 
\mu=0$, and for the temperature the BC is $\mathbf{n} \cdot \nabla \Theta=0$, where $\mathbf{n}$ is a unit 
vector normal to the boundary. For the 
velocity, we impose a tangential flow, $\mathbf{n} \cdot \mathbf{v} =0$, with a vanishing 
shear stress $(\mathbbm{1}-\mathbf{n} \mathbf{n})\cdot \tau=0$, where $\tau=(\nabla\mathbf{v}+\nabla\mathbf{v}^T)$ is the viscous stress 
tensor. 

The boundaries 
at $z=\pm l$ can be an inlet or outlet for the flow, depending on the colloid's 
translation direction relative to its cap. At the inlet, we have a 
critical mixture, $\varphi=0$, with a temperature $\Theta=0$ and a 
velocity $\mathbf{v}=-\mathbf{n} $. At the outlet, the mixture is freely 
advected, and we impose a vanishing  diffusive fluxes: 
$\mathbf{n} \cdot \nabla \mu=0$ and $\mathbf{n} \cdot \nabla \Theta=0$. We also impose a vanishing total
stress: $\mathbf{n}\cdot\left(p\mathbbm{1}+\Pi-\tau \right)=0$, where $\Pi$ is the Korteweg stress tensor~\cite{anderson1998}: $\Pi =\left[ \left(\varphi\partial f_0/\partial 
\varphi-f_0\right)-C|\nabla\varphi|^2/2-C\varphi\nabla^2\varphi\right] 
\mathbbm{1}+C\nabla\varphi\nabla\varphi$.

The remaining boundaries to be addressed are the two chemically distinct 
colloid surfaces. For the velocity, we impose a no-slip BC on both hemispheres, $\mathbf{v}=0$. 
The heat flux across the solid-fluid 
boundary is given by,
\begin{equation}
\label{eq:heatflux}
-\mathbf{n}
\cdot \nabla \Theta_{fluid} +\mathbf{n} \cdot 
K \nabla \Theta_{solid}=H(c)
\end{equation}
where $K=k_s/k_f$ is the conductivity contrast, $H$ 
is the Heaviside step function and $c=-\cos(\vartheta-\alpha)$, where $\vartheta$ is the polar angle relative to $\mathbf{e}$
and $\alpha = 0.05\pi$ determines the carbon coverage, as depicted in~Fig.~\ref{fig:supp}.

For the composition BC we use
\begin{eqnarray}
 \label{eq:bc_comp1}
 \mathbf{n} \cdot \nabla \mu&=&0~,\\
 \mathbf{n} \cdot \nabla \varphi &=& -\tan\left(\frac{\pi}{2}-\theta_i\right)\left|  
\nabla \varphi - \left(\mathbf{n} \cdot \nabla \varphi\right)\mathbf{n}\right|~.
 \label{eq:bc_comp2}
\end{eqnarray}
(\ref{eq:bc_comp1}) imposes no material flux at the boundary, 
while (\ref{eq:bc_comp2}) imposes the contact angle $\theta_i$, where $i=1,2$ 
denotes the capped and uncapped areas, respectively. This so-called 
geometric formulation of the wetting BC 
has proved useful in diffuse interface simulations 
of moving contact lines. It ensures that $\varphi$ is adjusted such 
the that $\theta_i$ is imposed at the surface and contours of $\varphi$ are 
tangent to the interface. The contact angles are 
related to the short-range interactions between the liquid and solid via: 
$\cos\theta_i=\sqrt{2}\gamma_i$ \cite{ding2007}, where $\gamma_i$ is the surface 
field in the linear surface free-energy density, $f_s^i=\gamma_i\varphi$. 
The value of contact angles in the experiments is unknown and 
we therefore use an indicative value of $\theta_1=\pi/4$ for the hydrophobic heated cap and 
$\theta_2=3\pi/4$ for the hydrophilic uncapped area. With this choice, $\varphi>0$ corresponds to a 
PnP-rich phase.

To obtain the swimming velocity, we use the following procedure.
For a given set of parameters, we relax the velocity and 
composition towards steady state, using as input the temperature field 
obtained from (\ref{eq:lap}), and two initial guesses of 
$U$. The numerical solution is obtained using the software COMSOL Multiphysics 
v4.4.  We then calculate the force $\mathbf{F}$ exerted on the particle 
by the fluid by applying the divergence theorem to (\ref{eq:ns_dl}): 
\begin{equation}
\mathbf{F}=-2\pi\int_{-1}^{1}{\rm d} c \left[p\mathbbm{1}+\Pi-\tau 
\right]\cdot \mathbf{n}~,
\label{eq:force}
\end{equation}
We verified that the calculation of $\mathbf{F}$ is independent of the domain size $l$. At steady-state, the colloid 
should be force-free. We therefore adjust $U$ iteratively using the secant 
method, repeating the numerical solution of the governing equations until 
$\mathbf{F}$ approaches zero with a relative error of less than 1$\%$.

\section*{Acknowledgements}

This work was supported by the German Research Foundation (DFG) through grant No. GO 2797/1-1 (J.R.G.S.), by the DFG
through the priority programme SPP 1726 on microswimmers (C.B.), by the ERC Advanced Grant ASCIR grant No. 693683 (C.B.), by a Netherlands Organisation for Scientific Research (NWO) 
VICI grant (R.v.R.) funded by the Dutch Ministry of Education, Culture and Science 
(OCW), and by the European Union's Horizon 2020 programme under the Marie 
Sk\l{}odowska-Curie grant agreement No. 656327 (S.S.). This work is part of the D-ITP 
consortium, a program of NWO funded by OCW.

\section*{Author contributions statement}

J.R.G.S., C.L., and C.B. conceived the experiments,  J.R.G.S., C.L., and P.R.B. conducted the experiments, S.S. and R.v.R. concieved the model and carried out the numerical simulations, J.R.G.S., C.L., and S.S. analysed the results.  All authors reviewed the manuscript. 

\section*{Additional information}

\textbf{Competing financial interests:} The authors declare no competing financial interests.

\end{document}